# Magneto-elastic switching of magnetostrictive nanomagnets with in-plane anisotropy: The effect of material defects

Md Ahsanul Abeed[1], Jayasimha Atulasimha[1,2] and Supriyo Bandyopadhyay[1,*]

[1]Department of Electrical and Computer Engineering
[2]Department of Mechanical and Nuclear Engineering
Virginia Commonwealth University, Richmond, VA 23284, USA

## ABSTRACT

We theoretically study the effect of a material defect (material void) on switching errors associated with magneto-elastic switching of magnetization in elliptical magnetostrictive nanomagnets having in-plane magnetic anisotropy. We find that the error probability increases significantly in the presence of the defect, indicating that magneto-elastic switching is particularly vulnerable to material imperfections. Curiously, there is a critical stress value that gives the lowest error probability in both defect-free and defective nanomagnets. The critical stress is much higher in defective nanomagnets than in defect-free ones. Since it is more difficult to generate the critical stress in small nanomagnets than in large nanomagnets (having the same energy barrier for thermal stability), it would be a challenge to downscale magneto-elastically switched nanomagnets in memory and other applications where reliable switching is required. This is likely to be further exacerbated by the presence of defects.

---

[*] Corresponding author. Email: sbandy@vcu.edu

## 1. Introduction

Magneto-elastically switched nanomagnets [1-17] have an important niche in memory and logic applications since magneto-elastic switching of nanomagnets is remarkably energy-efficient [18-22]. Recent experiments on magneto-elastic switching however has shown that the switching probability of magnetostrictive nanomagnets deposited on a piezoelectric substrate is relatively small when a voltage is applied to the piezoelectric to generate strain in the nanomagnets [23-27]. While there may be many reasons for the low switching rate, one obvious possibility is that material defects in the nanomagnets might impede switching. Here, we have studied the role that material defects play in causing switching failures by simulating the effect of a physical void, or "hole", in the nanomagnet (one type of material defect) on the switching probability in the presence of thermal noise.

The scenario we are studying is the magneto-elastic switching of the magnetization of a magnetostrictive nanomagnet in a dipole coupled pair of elliptical nanomagnets, one of which (left nanomagnet in Fig. 1) has higher eccentricity than the other. The left nanomagnet is the "hard" nanomagnet because of its large shape anisotropy. It remains magnetized in one direction along its major axis and is not affected by (any reasonable amount of) stress since the stress anisotropy energy cannot overcome the large shape anisotropy energy in this nanomagnet and make its magnetization rotate. The right nanomagnet is the "soft" nanomagnet whose magnetization can be affected by stress since it has a lower shape anisotropy energy barrier. The two nanomagnets are positioned such that the line joining their centers is collinear with their minor axes (hard axes). In this case, shape anisotropy would prefer that the magnetizations of the two nanomagnets lie along their major axes (easy axes) and dipole coupling between the two would prefer that the magnetizations of the two nanomagnets be mutually antiparallel.

Let us suppose that we have magnetized both nanomagnets in the same direction along their major axes with a strong magnetic field (Fig. 1a). This will place the soft nanomagnet in a metastable state. Its magnetization is aligned along the easy axis, but it is not the preferred orientation or lowest energy state since the magnetization is parallel (not antiparallel) to that of the hard nanomagnet. The shape anisotropy energy barrier in the soft nanomagnet (which is smaller than that in the hard nanomagnet but still much larger than the thermal energy *kT*) prevents its magnetization from flipping spontaneously and assuming the antiparallel orientation. Essentially, the energy barrier prevents the transition from the metastable state (parallel magnetizations) to the ground state (antiparallel magnetizations).

We can erode the shape anisotropy energy barrier by applying a *critical stress* that *just* erodes the barrier but does not invert it. The critical stress is the stress that makes the stress anisotropy energy equal to the energy barrier, and hence it is ideally defined by the relation $(3/2)\lambda_s \sigma_{crit} \Omega = E_b$ where $\lambda_s$ is the magnetostriction coefficient of the soft nanomagnet, $\Omega$ is the nanomagnet volume, $E_b$ is the shape anisotropy energy barrier and $\sigma_{crit}$ is the critical stress. The application of critical or super-critical stress will remove the energy barrier and allow the soft nanomagnet's magnetization to flip and assume the antiparallel orientation, thus transitioning the system from the metastable state to the ground state.

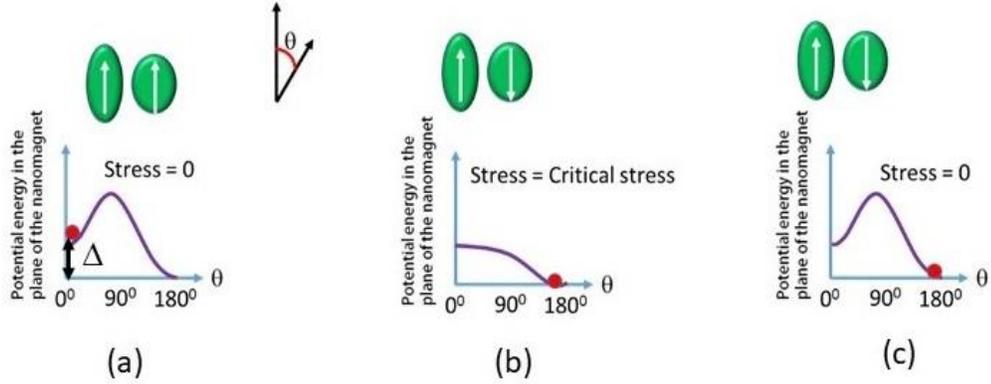

Fig. 1: (a) Two dipole-coupled elliptical magnetostrictive nanomagnets with in-plane anisotropy, the left more shape anisotropic than the right, are magnetized in the same direction along the easy axis with an external magnetic field. The potential profile of the right nanomagnet is shown under various stress conditions and the "ball" represents the state of the right nanomagnet determined by its magnetization orientation. The barrier in the potential profile is due to shape anisotropy and it is asymmetric due to dipole coupling with the left nanomagnet. When the right nanomagnet's magnetization is parallel to that of the left nanomagnet, the former nanomagnet is stuck in a metastable state and cannot transition to the ground state because of the intervening barrier; (b) the shape anisotropy energy barrier in the right nanomagnet is eroded (but not inverted) by applying *critical stress* (compressive or tensile) along the major axis. This allows the right nanomagnet to come out of the metastable state and reach the global energy minimum (ground state) and its magnetization flips to assume the antiparallel configuration; (c) the system remains in the global minimum state with the magnetizations antiparallel after stress is removed.

The application of "critical stress" actually results in more reliable switching than if "super-critical" (excess) stress is applied [28]. This is easy to understand. If excess stress is applied, then we would *invert* the potential barrier (instead of just eroding it), driving the system to the new energy minimum created at $\theta = 90°$ (magnetization pointing along the hard axis) [see Fig. 2]. Subsequent removal of stress would have more likely driven the system to the correct global minimum state at $\theta = 180°$ (because it is lower in energy than the local minimum state at $\theta = 0°$), but still with some significant probability of going to the incorrect local minimum state at $\theta = 0°$, as shown in Fig. 2, because the energy difference between the two states is only a few times $kT$ (for any reasonable dipole coupling strength). On the other hand, if critical stress is applied, then no energy minimum is ever created at $\theta = 90°$ (see Fig. 1) and hence the above problem does not arise. The energy minimum is created only at $\theta = 180°$ and therefore the probability of reaching the correct global minimum state at $\theta = 180°$ is much higher than that in the case of super-critical stress [28]. Thus, tailoring the stress to the critical value is important for reliable magneto-elastic switching.

The stress induced switching of the soft nanomagnet's magnetization to assume an orientation antiparallel to that of the hard nanomagnet implements the conditional dynamics of an inverter or a Boolean NOT gate. The magnetization orientation of the hard nanomagnet encodes the input bit and that of the soft nanomagnet the output bit. Since the ground state is the antiparallel configuration, the output bit is always the logic complement of the input bit as long as the system can migrate to the ground state. The stress acts as a "clock" to trigger the NOT operation by eroding the potential barrier and allowing the system to transition to the ground state. The operation of such a NOT gate, triggered by stress, or a surface acoustic wave, has been demonstrated experimentally [26, 29], but the switching probability as a function of stress magnitude has not been studied experimentally to test the hypothesis that critical stress produces the most reliable switching. The probable reason why it has not been studied is that the nanomagnets are usually stressed by applying a high voltage across an underlying piezoelectric substrate and piezoelectric materials cannot sustain too many cycles of stress ($< 10^7$ for PMN-PT) because of piezoelectric fatigue [30].

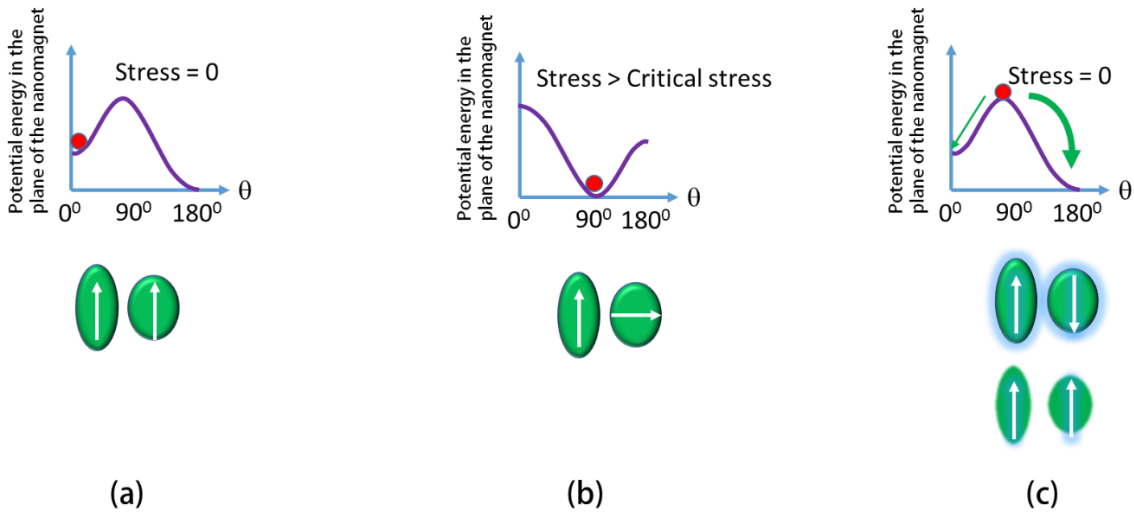

Fig. 2: (a) The right nanomagnet of Fig. 1 in the metastable state where its magnetization has been oriented parallel to that of the left nanomagnet by a magnetic field, (b) excess stress is applied to invert the shape anisotropy energy barrier and drive the magnetization of the right nanomagnet to the new energy minimum where it points along the minor axis of the ellipse (hard axis); (c) when stress is released, the magnetization of the right nanomagnet will prefer to align antiparallel to that of the left nanomagnet because of dipole coupling, but the probability of aligning parallel, albeit lower, is not small. Consequently, there is a significant probability of switching error if the nanomagnet is overstressed to "invert" the potential barrier as shown in Fig. 2(b).

Table 1 lists the critical stress values for a nanomagnet with a shape anisotropy energy barrier of 60 kT (room temperature) and volume 54,000 $nm^3$. We have assumed the bulk values of magnetostriction in amorphous materials. Materials with higher magnetostriction obviously require lower stress to switch.

Table 1: Critical stress values for an elliptical nanomagnet with a shape anisotropy energy barrier of 60 kT and volume 54,000 nm$^3$. The magnetostriction values are taken from references [31, 32].

| Material | Magnetostriction ($\lambda_s$) | Critical stress |
|---|---|---|
| Co | ~35 ppm | 88.5 MPa |
| Ni | ~35 ppm | 88.5 MPa |
| Terfenol-D | ~600 ppm | 5.16 MPa |

## 2. Understanding the error probability under critical stress

Consider the case of the two dipole-coupled elliptical nanomagnets discussed in the previous section. They act as a "stress-clocked" NOT gate. We will discuss error issues in this magneto-elastically triggered NOT gate under critical stress. Because the potential profile of the soft nanomagnet under critical stress will look like that in Fig. 3(a), we might naively assume that the probability of the soft nanomagnet remaining stuck in the original metastable state (failure to switch) would be ~ $e^{-\Delta/kT}$, where $\Delta$ is the energy difference between the metastable state and the ground state, as shown in Fig. 3(a). In reality, this is not true since the Boltzmann formulation can only apply in equilibrium and the system is not in equilibrium when it is switching.

It is perhaps also important at this point to make a distinction between static and dynamic error probabilities. The static probability is determined by the energy barrier separating the two stable magnetization states; it is the probability that the magnetization will spontaneously flip between these two states. Since this can happen only by thermionically crossing the barrier, the static error probability is approximately $e^{-E_b/kT}$. Thus, the barrier $E_b$ determines the thermal stability. The dynamic error, on the other hand, occurs during intentional switching when the barrier has been lowered by some external agent (stress, magnetic field, etc.). The dynamic error probability does not depend on the barrier height $E_b$ since that is eroded during switching. It is this quantity that we address in this paper.

For very slow switching, local equilibrium may be restored at every switching time step, and a Boltzmann-type analysis would be approximately valid. Since Boltzmann statistics is obeyed only in equilibrium (and not in a system driven far out of equilibrium), it is important for the switching to be slow enough that the magnetization distribution has time to relax to the equilibrium distribution at the given stress level. The magnetization distribution $\rho(\theta)$ is defined as the distribution of the probability that the macro-spin magnetization component in the plane of the nanomagnet is oriented along a particular direction denoted by the polar angle $\theta$.

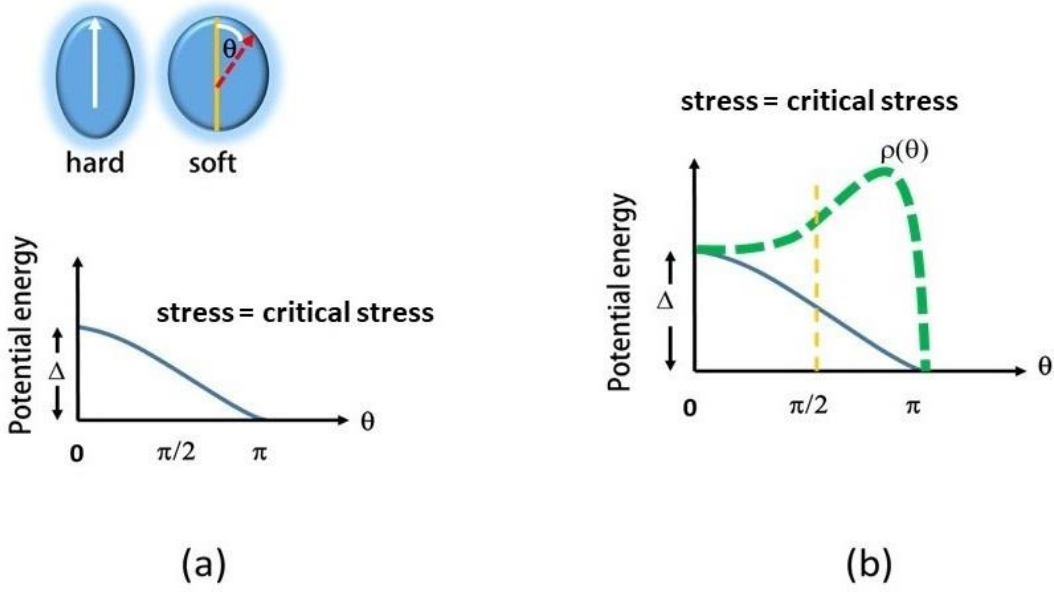

Fig. 3: (a) The potential energy profile of the soft nanomagnet dipole coupled to a hard nanomagnet (potential energy versus polar angle of the magnetization vector) when the nanomagnet is critically stressed; (b) the probability distribution of the magnetization orientation under critical stress, just before stress removal.

In Fig. 3(b), we qualitatively illustrate the probability distribution $\rho(\theta)$ of the magnetization vector of the output nanomagnet as a function of the polar angle $\theta$ in the presence of critical stress, just before the removal of the stress. This is the distribution of the magnetization orientation just before stress removal. When stress is removed (abruptly), the shape anisotropy energy barrier is immediately restored at $\theta = \pi/2$ and hence the area under the distribution curve to the left of the vertical dashed line at $\theta = \pi/2$ will be the probability that the magnetization state is trapped to the left of the potential barrier and will fail to switch. Therefore, the error probability will be

$$p = \frac{\int_0^{\pi/2} \rho(\theta)\,d\theta}{\int_0^{\pi} \rho(\theta)\,d\theta} \ . \tag{1}$$

The more skewed the probability distribution is to the right of the vertical dashed line, the lower is the error probability.

Ref. [33] studied the special situation when

$$\frac{V(\theta)-V(\pi)}{V(0)-V(\pi)} = \frac{V(\theta)-V(\pi)}{\Delta} = \frac{\pi-\theta}{\pi} \ , \tag{2}$$

where $V(\theta)$ is the potential energy of the right nanomagnet when its magnetization angle is $\theta$. Equation (2) would represent the situation when $V(\theta)$ is proportional $-\theta$, which means that the potential profile in Fig. 3(a) is linear. Assuming that local equilibrium prevails before stress removal, ref. [32] used the following Boltzmann-type expression for $\rho(\theta)$:

$$\rho(\theta) = Ae^{-(V(\theta)-V(\pi))/kT} = Ae^{-(\pi-\theta)\Delta/\pi kT} , \quad (3)$$

where $A$ is found from the normalization condition $\int_0^\pi \rho(\theta)d\theta = 1$, yielding

$$A = \frac{\Delta}{kT}\frac{1}{1-e^{-\Delta/kT}} .$$

Using this result in Equation (1), one gets

$$p = e^{-\Delta/4kT}\frac{\sinh\left(\dfrac{\Delta}{4kT}\right)}{\sinh\left(\dfrac{\Delta}{2kT}\right)} \approx e^{-\Delta/2kT} \quad [\text{if } \Delta \gg 2kT] . \quad (4)$$

According to the above relation, the energy dissipated in the switching can be related to the error probability as $\Delta = 2kT\ln(1/p)$. All these results are, of course, approximate and can hold true only when sufficient time is allowed for the probability distribution to reach local equilibrium with the potential profile after the barrier is lowered. This will ensure that Equation (3) is obeyed. After that, the barrier is raised back abruptly, so the error probability is the area under the probability distribution curve to the left of the barrier peak. As such, these distributions only provide a qualitative understanding of the switching error. In reality, the 3-dimensional distribution of magnetization orientation (note: the azimuthal angle, Φ, has not be considered in the above derivation) produces a vastly more complex magnetization dynamics when the barrier is restored.

### 3. Magneto-elastic switching in a nanomagnet with a material defect

When a material defect is present in a nanomagnet, the above analytical formulations become too inaccurate and the switching error probability has to be found from numerical simulations. Here, we study magneto-elastic switching in an elliptical magnetostrictive nanomagnet containing a material defect, e.g. a material void, implemented as a physical "hole". We will assume that initially a nanomagnet is in one of its stable orientations along the major axis (easy axis) and a magnetic field is applied in the opposite direction to make its magnetization flip. The magnetic field however is too weak to overcome the shape anisotropy energy barrier separating the two stable orientations along the easy axis and hence uniaxial stress (of the correct sign) is applied along the major axis of the ellipse to erode or invert the energy barrier and allow the nanomagnet to switch its magnetization to point in the direction of the magnetic field.

The magnetic field can be viewed as the dipole coupling field due to the presence of a nearby elliptical nanomagnet whose major axis is parallel to the major axis of the test nanomagnet and the line joining their centers is collinear with their minor axes (i.e. the NOT gate). We will study the switching of the magnetization of the test nanomagnet (with a hole in its center) due to the combined effect of the bias magnetic field and stress in the presence of thermal noise using the micromagnetic simulator MuMax3 [34]. The procedure for incorporating stress and thermal noise in the simulations has been described in ref. [35] and will not be repeated here.

The switching error probability is found as follows: At time $t = 0$, the magnetization points antiparallel to the bias magnetic field. We then turn on both the magnetic field and stress abruptly at time $t = 0$ and generate 1000 switching trajectories in MuMax3 under the combined effect of the bias magnetic field and stress. The simulations are carried out until each trajectory reaches a steady state and the magnetization aligns either parallel or antiparallel to the bias magnetic field. The time taken to reach steady state is ~1 ns. The fraction of trajectories that end up with the magnetization antiparallel to the magnetic field (fails to switch) is the switching error probability. The time step used in the simulations is 10 fs and the spatial resolution is 2 nm.

The test nanomagnet is assumed to be made of Terfenol-D and has major axis dimension 100 nm, minor axis dimension 90 nm and thickness 6 nm. The bias magnetic field has a flux density of 3 mT (which would be a typical value for dipole coupling field for nanomagnets of these dimensions; recall that we are simulating the NOT dynamics of a dipole-coupled pair and the nanomagnet hosting the output bit is subjected to the dipole coupling field due to the nanomagnet hosting the input bit) and uniaxial compressive stress is applied along the major axis to erode or invert the energy barrier in the nanomagnet. The critical stress for this nanomagnet is 2.26 MPa in the absence of the magnetic field (shape anisotropy barrier = 26.3 kT). The magnetic field will lower it to 0.78 MPa by lowering the barrier to ~9 kT.

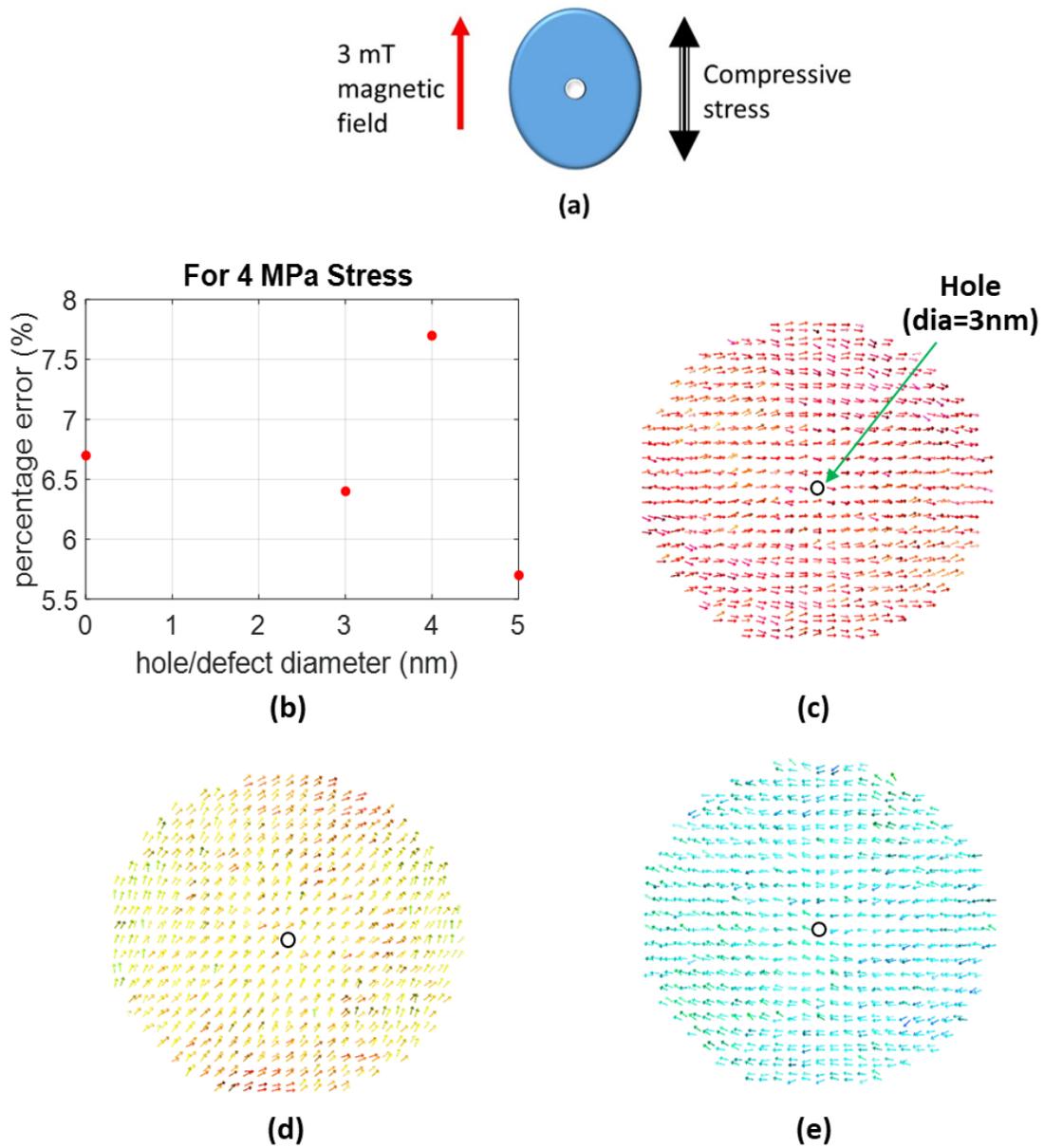

Fig. 4: (a) Nanomagnet with a hole at the center. A magnetic field of 3 mT is applied along the major axis to flip the magnetization and uniaxial compressive stress is also applied along the major axis to aid the flipping. (b) Switching error probability as a function of the hole diameter when the applied stress is 4 MPa; (c) Magnetization texture in the initial state (with a 3 nm diameter hole in the center); (d) Magnetization texture under a stress of 4 MPa; (e) Magnetization texture after release of stress.

We introduce a material defect in the form of a circular hole in the elliptical nanomagnet's center. The diameter of this hole is varied to study the switching error probability as a function of the hole diameter under three different stress magnitudes (sub-critical, near-critical and super-critical). The critical stress will be larger in the defective nanomagnet than in the defect-free nanomagnet since the energy barrier separating the two stable magnetization orientations is made larger by the hole (recall that the critical stress is proportional to the energy barrier). This happens because the energy barrier is due to shape anisotropy. The introduction of the hole makes the ellipse into an annulus and increases the effective shape anisotropy. Therefore, we expect that the hole will increase the energy barrier and hence the critical stress. There is no direct recipe for calculating the energy barrier in a nanomagnet with a hole and hence a more rigorous description is left for future work.

We note that while the simulator MuMax3 accounts for demagnetization energy changing with the introduction of a hole, it still assumes uniform stress distribution and does not account for spatial variation of stress in the vicinity of the hole. The spatial variation is ignored to keep the computations tractable.

The results of the simulations are shown in Figs. 4-7. In each of these figures, we show the switching error probability as a function of the hole diameter (0 nm, 3 nm, 4 nm and 5 nm) for stresses of 4 MPa (sub-critical), 8 MPa (near critical), 16 MPa (near critical) and 50 MPa (super-critical, excess stress). The sub-critical stress is insufficient to make the stress anisotropy energy erode the energy barrier separating the two stable magnetization orientations in the presence of the 3 mT magnetic field, and hence, expectedly, the switching probability is small causing the error probability to be very large for sub-critical stress. The 8 MPa and 16 MPa stresses cause the minimum error probabilities – with or without the hole. Without the hole, none out of 1000 simulated switching trajectories failed to switch when 8 MPa or 16 MPa was applied, showing that the switching error probability is less than 0.1% for these two stress values in a defect-free nanomagnet. For a 3 nm diameter hole, the error probability is still less than 0.1% for 8 MPa stress (none of the 1000 trajectories failed to switch) and becomes 0.1% for 16 MPa stress (one out of 1,000 trajectories failed to switch). When the stress is increased to 50 MPa, the error probability increases for all hole diameters, showing that there is an optimum stress (for lowest switching error probability) even when there is a material defect (e.g. a physical hole) in the nanomagnet, except it is much higher in a defective nanomagnet than in a defect-free nanomagnet (recall that the calculated critical stress in a defect-free nanomagnet is 0.78 MPa).

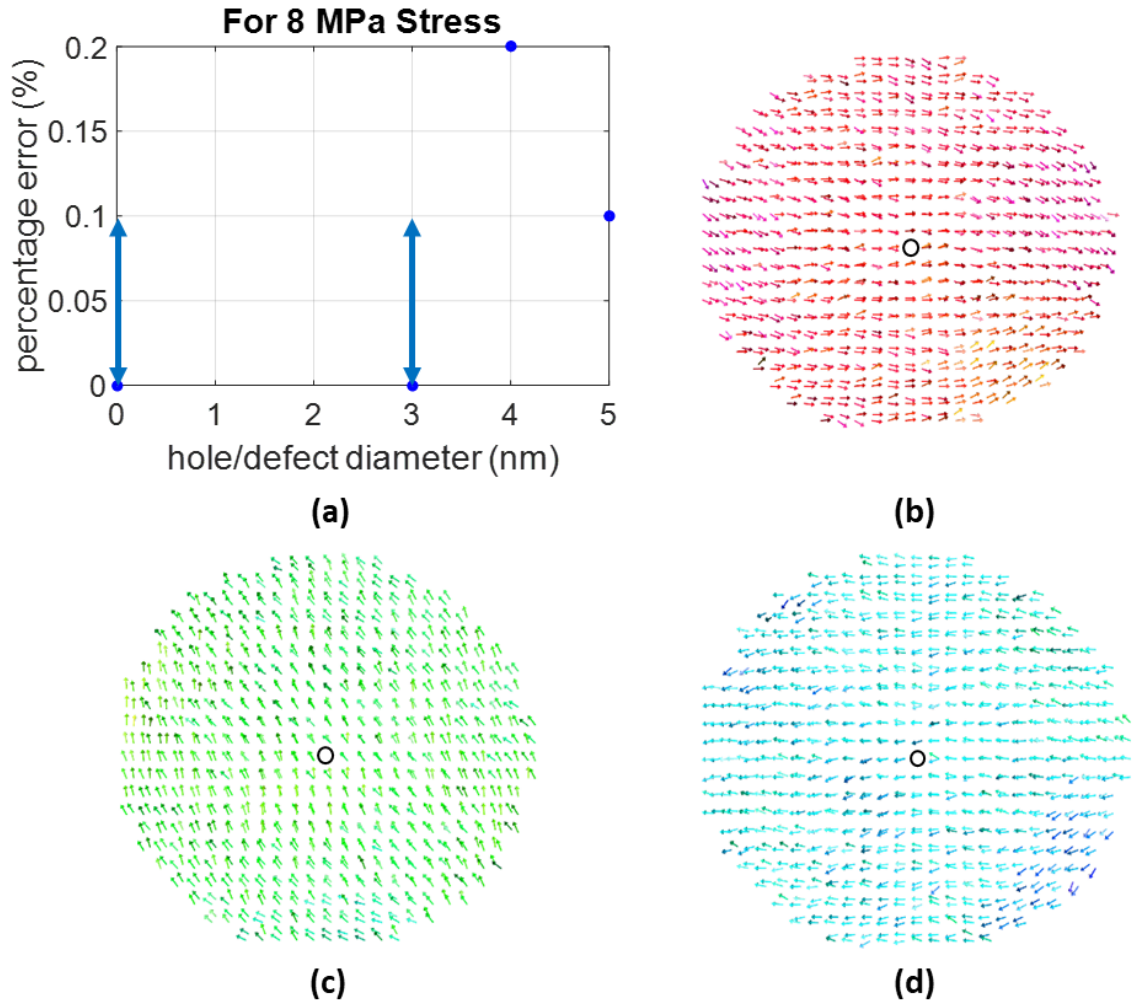

Fig. 5: (a) Switching error probability as a function of the hole diameter when the applied stress is 8 MPa; (b) Magnetization texture in the initial state (with a 3 nm diameter hole in the center); (c) Magnetization texture under a stress of 8 MPa; (d) Magnetization texture after release of stress.

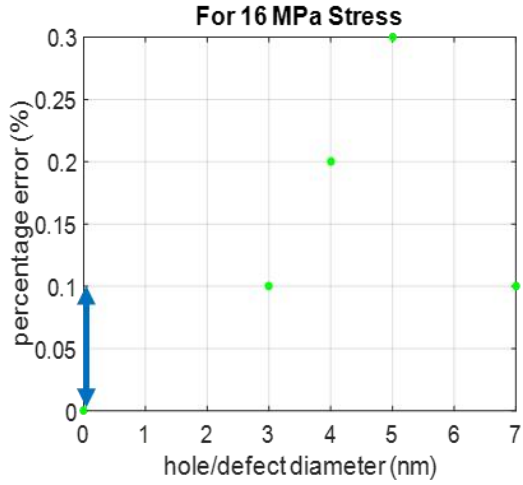 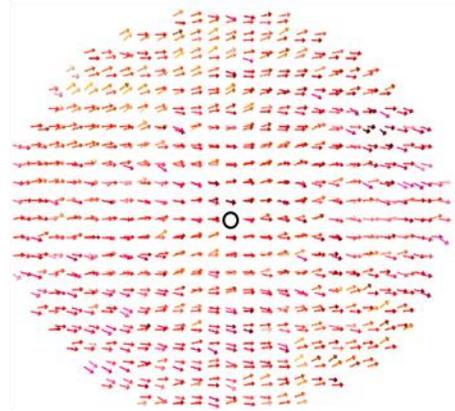
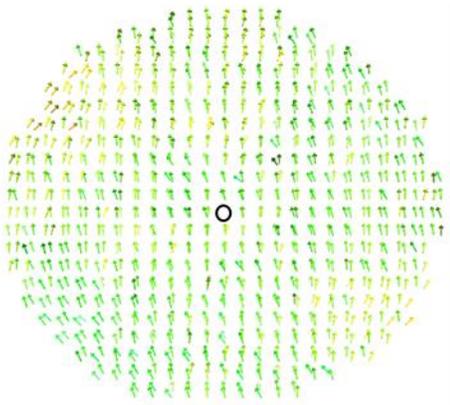 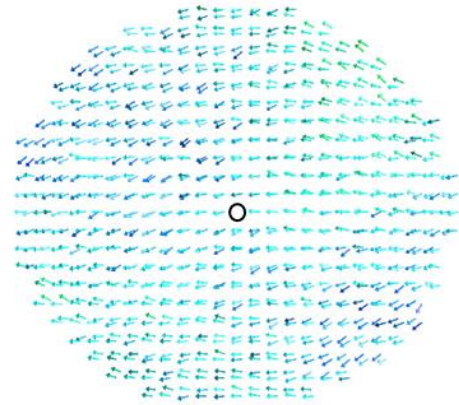

Fig. 6: (a) Switching error probability as a function of the hole diameter when the applied stress is 16 MPa; (b) Magnetization texture in the initial state (with a 3 nm diameter hole in the center); (c) Magnetization texture under a stress of 16 MPa; (d) Magnetization texture after release of stress.

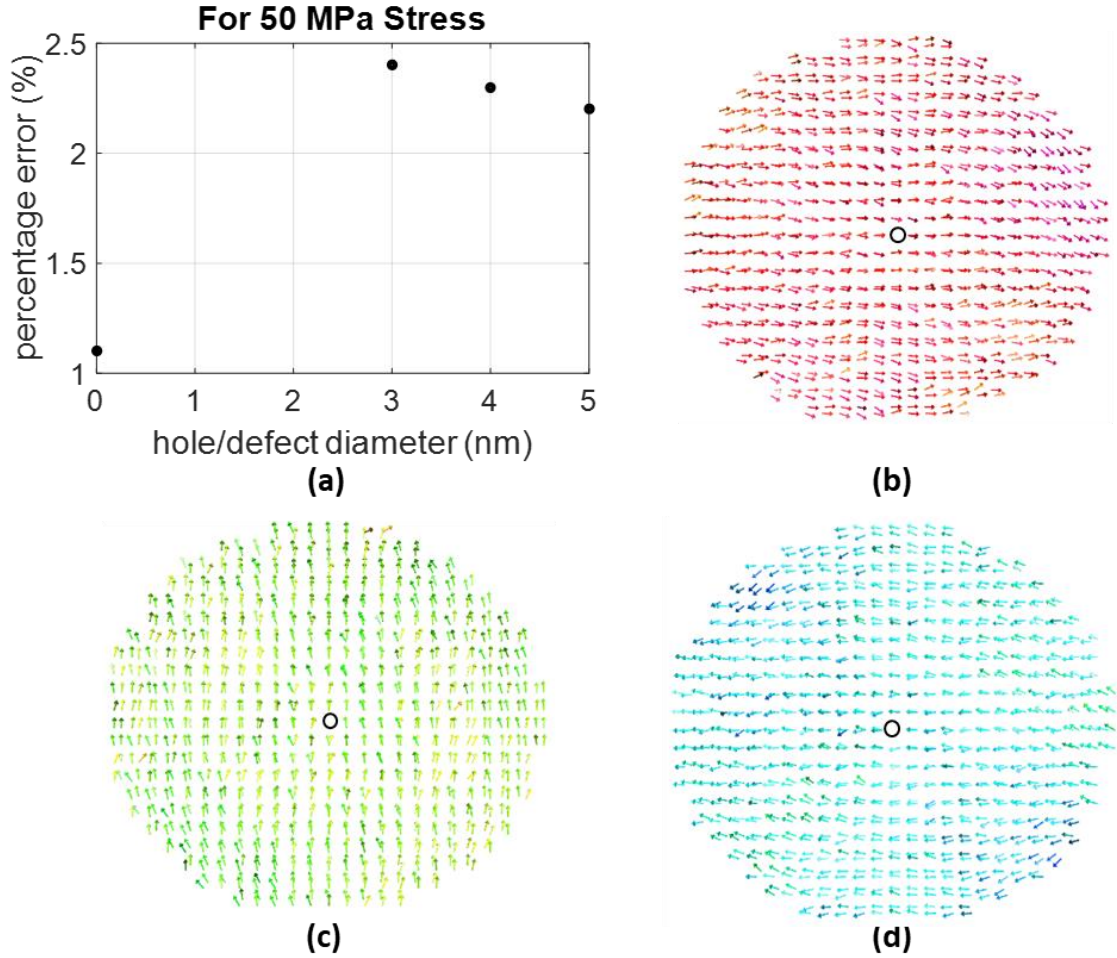

Fig. 7: (a) Switching error probability as a function of the hole diameter when the applied stress is 50 MPa; (b) Magnetization texture in the initial state (with a 3 nm diameter hole in the center); (c) Magnetization texture under a stress of 50 MPa; (d) Magnetization texture after release of stress.

It is obvious from Figs. 4-7 that the switching error probability is sensitive to material defects since the probability always increases when there is a hole, provided the applied stress is either critical or super-critical (enough to erode or invert the barrier). Thus, magneto-elastic switching is not robust against material defects. These conclusions reinforce another recent work that uses different defect configuration to explain the critical role of defects in impeding magnetization control through the inverse magneto elastic (Villari) effect [36]. All this tells us that excellent material quality is required for *reliable* magneto-elastic switches.

An interesting observation is that for a given stress magnitude, the error probability does not increase monotonically with the diameter of the hole. This is not surprising. Different hole sizes produce different microscopic magnetization distribution within the nanomagnet at a given stress. It is entirely possible that the distribution corresponding to a smaller hole is more vulnerable to switching error than that corresponding to a larger hole. Thus, ensuring that defect size remains small does not guarantee that the error probability will remain small as well when stress of a fixed magnitude is applied. There is no specific trend in the dependence of the error probability on hole diameter because the microscopic distributions of the magnetizations within the nanomagnet do not have a specific or predictable dependence on hole diameter at any given stress value.

In Fig. 8, we plot the error percentage as a function of stress for a 3-nm diameter hole located at the center of the nanomagnet. There is clearly a region of minimum error probability which establishes that there is a critical stress value for maximum error resilience. The same trend holds for other hole sizes (4 nm, 5 nm diameter) as well.

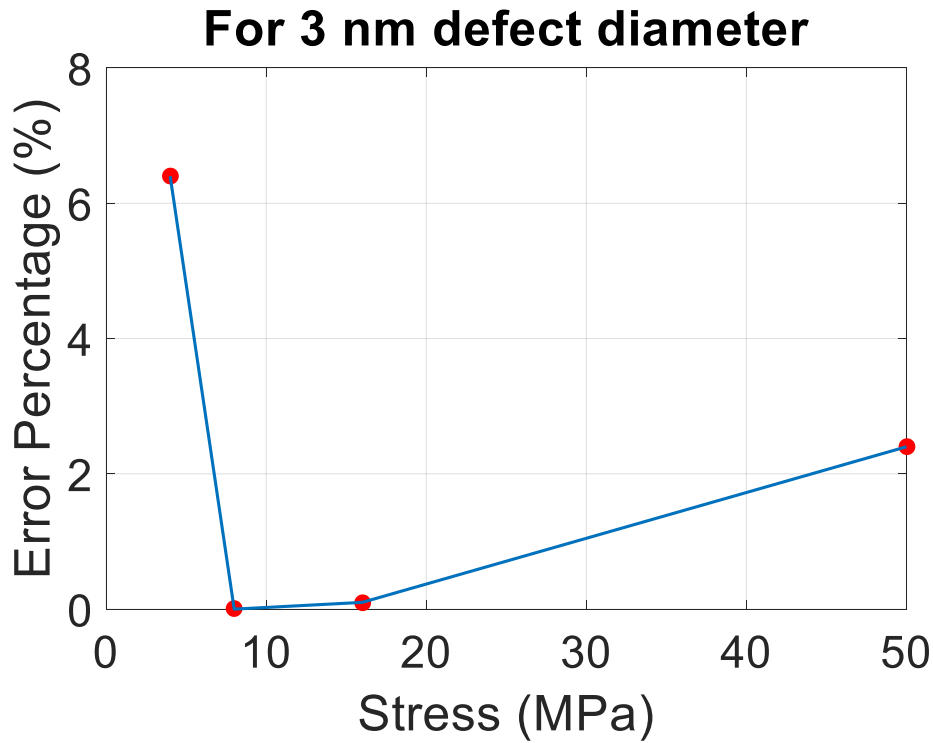

Fig. 8: Error probability versus stress for a 3 nm diameter hole. There is a region of stress that results in the lowest error probability and this region corresponds to the critical stress.

We also investigated if the location of the hole has any significant effect on the error probability. In Figs. 4-7, we had placed the hole in the center of the nanomagnet. We studied the effect of off-center holes in the case when the hole diameter is 3 nm and the applied stress is 4 MPa. In this case, the switching error probability is 6.4% when the hole is in the nanomagnet's center, and it changes to 7.2% when the hole is shifted along the minor axis by 20 nm to the right. It changes to 7.4% when the hole is shifted by 20 nm along the major axis to the top, and to 6.2% when the hole is shifted to a location (-10 nm, -10 nm), assuming that the center is at the origin. The magnetization distributions for these cases are shown in Fig. 9.

The reason hole location affects the error probability is because the latter is determined by the micromagnetic distributions within the nanomagnet at any given stress. Insofar as these distributions are altered by changing the location of the hole, it stands to reason that the hole position will affect the error probability.

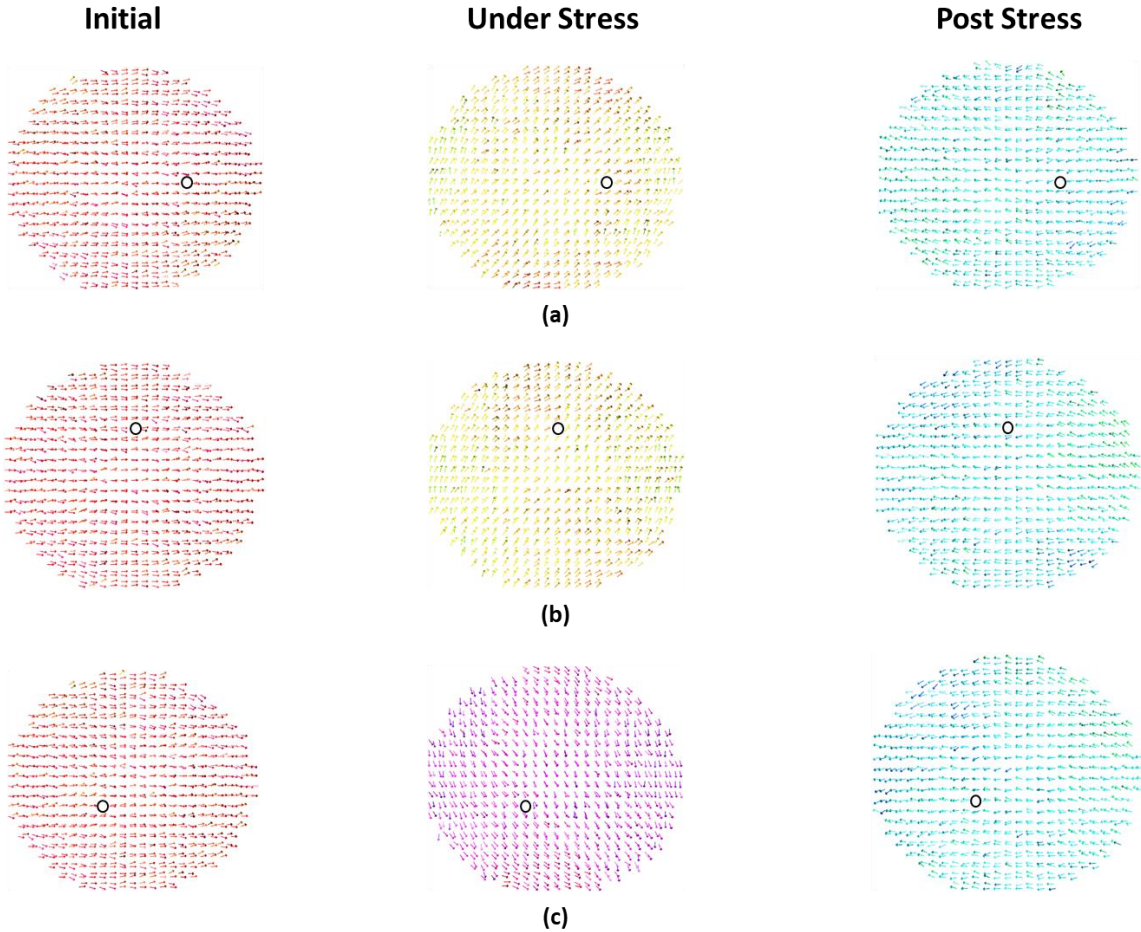

Fig. 9: Micromagnetic distributions when the hole is located: (a) 20 nm to the right of the center on the minor axis, (b) 20 nm above the center on the major axis, and (c) at the coordinate (-10 nm, -10 nm) assuming the center is the origin. The hole diameter is 3 nm and the applied stress is 4 MPa.

## 4. Conclusion

In this article, we have shown that magneto-elastic switching is vulnerable to material voids; defects exacerbate switching failures in the presence of thermal noise. We also showed that there exists a critical stress that results in the most reliable magneto-elastic switching in both defective and defect-free nanomagnets. If we assume, reasonably, that the maximum strain that can be generated in a nanomagnet is ~500 ppm, then based on the Young's moduli of Co, Ni and Terfenol-D, the maximum stresses that can be generated in these magnetostrictive materials are ~100 MPa for Co and Ni, and ~50 MPa for Terfenol-D. Comparing these figures with the critical stress values listed in Table 1, one finds that it should be possible to generate critical stresses in generic defect-free nanomagnets (with a shape anisotropy energy barrier of 60 *kT*) to obtain optimally reliable switching in nanomagnets of volume ~ 54,000 nm$^3$. There are however two primary concerns. The first is that defects and fabrication imperfections can make the switching much less reliable than what we might infer for defect-free nanomagnets because the critical stress value increases in the presence of defects. The second is that as one downscales nanomagnets to lateral dimensions of ~ 20 nm or less and thickness ~ 2 nm (which would be desirable for high-density non-volatile memory applications) while maintaining a fixed energy barrier $E_b$ for thermal stability, the critical stress (inversely proportional to nanomagnet volume) will become correspondingly higher and may be difficult, or impossible, to generate within the nanomagnet. Since the critical stress is higher in nanomagnets with a defect, the problem is more severe in the presence of defects. Thus, reliable magneto-elastic switching of nanomagnets with in-plane magnetic anisotropy may not have much tolerance for defects and may also suffer from serious scaling limitations.

Finally, one question that naturally arises is whether magneto-elastic switching is any more vulnerable to material defects of the type studied here than other magnet switching mechanisms like spin-transfer-torque (STT) [37]. While we have not carried out a rigorous quantitative comparison, it is likely that magneto-elastic coupling is more vulnerable to defect-induced errors than STT since there is no concept of barrier lowering to switch in the case of STT. In fact, there is no potential barrier picture associated with STT because the Slonczewski torque [37] is non-conservative, which precludes a potential barrier description of the switching. Another obvious difference is that the switching error probability can be always reduced arbitrarily in STT by increasing the STT current density. This is true even in the presence of a hole since STT works by transferring spin angular momentum from electrons in a spin polarized current to resident electrons in a nanomagnet (the hole may affect the STT current density required to attain a certain error probability). If we increase the STT current density, there will be more transfer of spin angular momentum per unit area of the nanomagnet, which will reduce the switching error probability. However, that trend does not exist in magneto-elastic switching since we have shown that we cannot decrease the error probability arbitrarily by increasing stress; in fact, too much stress (super-critical stress) is counter-productive. Thus, magneto-elastic switching has its own special features that make it distinct from the more popular switching mechanism of STT.

**Acknowledgement:** This work was supported by the US National Science Foundation (NSF) under grant ECCS 1609303. J.A is also supported by NSF CAREER grant CCF 1253370.